\documentclass[%
reprint,
superscriptaddress,
amsmath,amssymb,amsthm,mathrsfs,
aps,
prl,
]{revtex4-1}

\usepackage{graphicx}
\usepackage{dcolumn}
\usepackage{bm}
\usepackage[usenames]{color}
\usepackage[ 
margin=0.75in,
]{geometry}
\usepackage[normalem]{ulem}

\def\TN{$T_\mathrm{N}$}

\def\CM{$C_M$}

\def\muSR{$\mu$SR}
\def\muB{$\mu_{\mathrm{B}}$}

\begin{document}
	
\title{
	Spin dynamics and a nearly continuous magnetic phase transition in an entropy-stabilized oxide antiferromagnet
}
	
	\author{Benjamin A. Frandsen}
	\email{benfrandsen@byu.edu}
	\affiliation{ %
		Department of Physics and Astronomy, Brigham Young University, Provo, Utah 84602, USA.
	} %

	\author{K. Alec Petersen}
	\affiliation{ %
		Department of Physics and Astronomy, Brigham Young University, Provo, Utah 84602, USA.
	} %
	
	\author{Nicolas A. Ducharme}
	\affiliation{ %
		Department of Physics and Astronomy, Brigham Young University, Provo, Utah 84602, USA.
	} %
	
	\author{Alexander G. Shaw}
	\affiliation{ %
		Department of Physics and Astronomy, Brigham Young University, Provo, Utah 84602, USA.
	} %

	\author{Ethan J. Gibson}
	\affiliation{ %
		Department of Physics and Astronomy, Brigham Young University, Provo, Utah 84602, USA.
	} %

	\author{Barry Winn}
	\affiliation{ %
	Neutron Scattering Division, Oak Ridge National Laboratory, Oak Ridge, Tennessee 37831, USA.
	} %
	
	\author{Jiaqiang Yan}
	\affiliation{ %
		Materials Science \& Technology Division, Oak Ridge National Laboratory, Oak Ridge, Tennessee 37831, USA.
	} %
	
	\author{Junjie Zhang}
	\affiliation{ %
		Materials Science \& Technology Division, Oak Ridge National Laboratory, Oak Ridge, Tennessee 37831, USA.
	} %
	\affiliation{ %
		Institute of Crystal Materials \& State Key Laboratory of Crystal Materials, Shandong University, Jinan, Shandong 250100, China.
	} %

	\author{Michael E. Manley}
	\affiliation{ %
		Materials Science \& Technology Division, Oak Ridge National Laboratory, Oak Ridge, Tennessee 37831, USA.
	} %
	
	\author{Rapha\"el P. Hermann}
	\email{hermannrp@ornl.gov}
	\affiliation{ %
		Materials Science \& Technology Division, Oak Ridge National Laboratory, Oak Ridge, Tennessee 37831, USA.
	} %

\begin{abstract}
The magnetic order and the spin dynamics in the antiferromagnetic entropy-stabilized oxide (Mg$_{0.2}$Co$_{0.2}$Ni$_{0.2}$Cu$_{0.2}$Zn$_{0.2}$)O (MgO-ESO) have been studied using muon spin relaxation (\muSR) and inelastic neutron scattering. We find that antiferromagnetic order develops gradually in the sample volume as it is cooled below 140~K, becoming fully ordered around 100~K. The spin dynamics show a critical slowing down in the vicinity of the transition, and the magnetic order parameter grows continuously in the ordered state. These results indicate that the antiferromagnetic transition is continuous but proceeds with a Gaussian distribution of ordering temperatures. The magnetic contribution to the specific heat determined from inelastic neutron scattering likewise shows a broad feature centered around 120~K. High-resolution inelastic neutron scattering further reveals an initially gapped spectrum at low temperature which sees an increase in a quasielastic contribution upon heating until the ordering temperature.
\end{abstract}
	
\maketitle

\section{Introduction}	
In entropy-stabilized oxides (ESOs), a large number of cations (typically 5 or more) are combined in equimolar proportions to form new material phases~\cite{mirac;amt17,drago;s19}. ESOs are stabilized by the large configurational entropy resulting from so many cations distributed randomly and uniformly through the crystal lattice, allowing the formation of distinct phases not accessible by enthalpic considerations alone. Since the first discovery of an ESO system in 2015~\cite{rost;nc15}, numerous additional ESOs have been reported, many with outstanding mechanical, dielectric, optical, and electrochemical properties~\cite{mirac;amt17,sarka;am19}. This large class of materials therefore presents rich opportunities for establishing the fundamental physics of entropically stabilized systems and exploring potential technological applications. 
	
Until recently, the magnetic properties of ESOs had been only minimally studied. However, the discovery of long-range antiferromagnetic (AF) order in the rocksalt-type material (Mg$_{0.2}$Co$_{0.2}$Ni$_{0.2}$Cu$_{0.2}$Zn$_{0.2}$)O (hereafter called MgO-ESO) was a watershed moment for magnetic studies of ESOs, as it was the first observation of long-range magnetic order in a bulk ESO material~\cite{jimen;apl19,zhang;cm19}. Magnetism is now being actively explored in other ESOs as well, such as perovskite-type systems~\cite{witte;prm19}.
	
Through a combination of magnetometry and x-ray/neutron diffraction, MgO-ESO has been found to form a homogeneous fcc structure with a random and uniform distribution of the 5 distinct divalent cations~\cite{jimen;apl19,zhang;cm19}, with long-range AF order developing below \TN~$\sim$~113~K. This AF state is characterized by a propagation vector of (1/2, 1/2, 1/2) and consists of ferromagnetic alignment within (111) planes and antiferromagnetic alignment between these planes, similar to the magnetic structure of the simple binary oxides CoO and NiO~\cite{roth;pr58}. The existence of this AF state is remarkable in light of the extreme chemical disorder in MgO-ESO, including the presence of nonmagnetic cations Mg$^{2+}$ and Zn$^{2+}$. Even the magnetic cations are quite different from each other, with Cu$^{2+}$ (d$^{9}$), Ni$^{2+}$ (d$^{8}$), and Co$^{2+}$ (d$^{7}$) having quantum spin numbers $S$ = 1/2, 1, and 3/2, respectively. The establishment of such a simple AF structure amidst such magnetic complexity is quite intriguing. 
	
Important questions about this magnetic state remain open. For example, the magnetic susceptibility shows a clear kink at 113~K, yet the magnetic Bragg peak intensity from neutron diffraction begins growing at 140~K, while the specific heat shows no obvious signature of a transition at any temperature~\cite{zhang;cm19}. The origin of these discrepancies in temperature and detectability of the transition is not clear, although the presence of strong magnetic fluctuations from inelastic neutron scattering data is thought to play a role~\cite{zhang;cm19}. It is also not understood why the ordered magnetic moment determined from fits to the neutron diffraction data is 1.4~\muB, substantially smaller than the expected average of 2~\muB\ for fully ordered Cu$^{2+}$, Ni$^{2+}$, and Co$^{2+}$ moments. This reduced moment suggests that either the ordered volume fraction does not reach 100\% or that the magnetic moment is partially quenched~\cite{zhang;cm19}. Finally, whether the AF transition is first-order or continuous remains unknown.
	
Here, we report muon spin relaxation (\muSR) and inelastic neutron scattering (INS) measurements that address these open questions about the AF state in MgO-ESO. Using the sensitivity of \muSR\ to the magnetically ordered volume fraction, we determine that the AF order develops gradually throughout the sample between 100 and 140~K, with the midpoint of the transition occurring at 121.5~$\pm$~0.5~K. Below 100~K, the AF order extends through the full sample volume. The magnetic order parameter as measured by \muSR\ rises continuously from zero as the temperature is lowered through the transition, and critical spin dynamics are observed in the vicinity of the transition, demonstrating that the AF transition is nearly continuous. The only deviation from ideal continuous behavior appears to be the coexistence of the paramagnetic and AF phases during the finite temperature window in which the transition occurs. Therefore, one may consider the AF transition to be continuous, but with a distribution of N\'eel temperatures. The INS data are used to isolate the magnetic contribution to the specific heat, which is found to show a broad peak centered at approximately 120~K, consistent with the gradual transition revealed by \muSR. A magnon gap of $\sim$7 meV is observed at low temperature and is gradually covered by quasielastic spin relaxation scattering.

\section{Experimental Methods}		
The sample studied here was from the same batch used for the x-ray diffraction, neutron scattering, magnetometry, and heat capacity measurements reported in Ref.~\onlinecite{zhang;cm19}. The \muSR\ studies were conducted at TRIUMF in Vancouver, Canada using the LAMPF (Los Alamos Meson Physics Facility) spectrometer. In a \muSR\ experiment, spin-polarized positive muons are implanted one at a time in the sample. The spin of the muon undergoes Larmor precession around the local magnetic field, which is the vector sum of the internal field originating from the intrinsic magnetic moments and any externally applied field. After a mean lifetime of 2.2~$\mu$s, the muon decays into a positron and two neutrinos, with the positron being emitted preferentially in the direction of the muon spin at the moment of decay. Detectors near the sample position record the positron events. The normalized difference in positron events (i.e., the asymmetry) between a pair of detectors placed on opposite sides of the sample is recorded as a function of time. This asymmetry is proportional to the projection of the muon spin ensemble polarization along the axis defined by the detector pair. Analysis of the time-dependent \muSR\ asymmetry can be used to probe the internal magnetic field distribution in the sample~\cite{yaoua;b;msr11}. The temperature of the sample was controlled using a helium gas flow cryostat. Fits to the \muSR\ spectra were performed using the program MUSRFIT~\cite{suter;physproc12}. Inelastic neutron scattering data, the temperature dependence of the lattice constant, and the magnetic susceptibility data from Ref.~\onlinecite{zhang;cm19} were utilized to obtain thermodynamic quantities. Additional high-resolution inelastic neutron scattering data were collected at the BL-14B HYSPEC beamline of the Spallation Neutron Source between 5 and 300~K on 8.8~g of powder, the same sample as in Ref.~\onlinecite{zhang;cm19}. An incident energy of 15~meV and chopper frequency of 180~Hz were chosen, with the detector vessel positioned at -33 degree.
	
\section{Results}	
\subsection{Muon Spin Relaxation}
\begin{figure}
	\includegraphics[width=75mm]{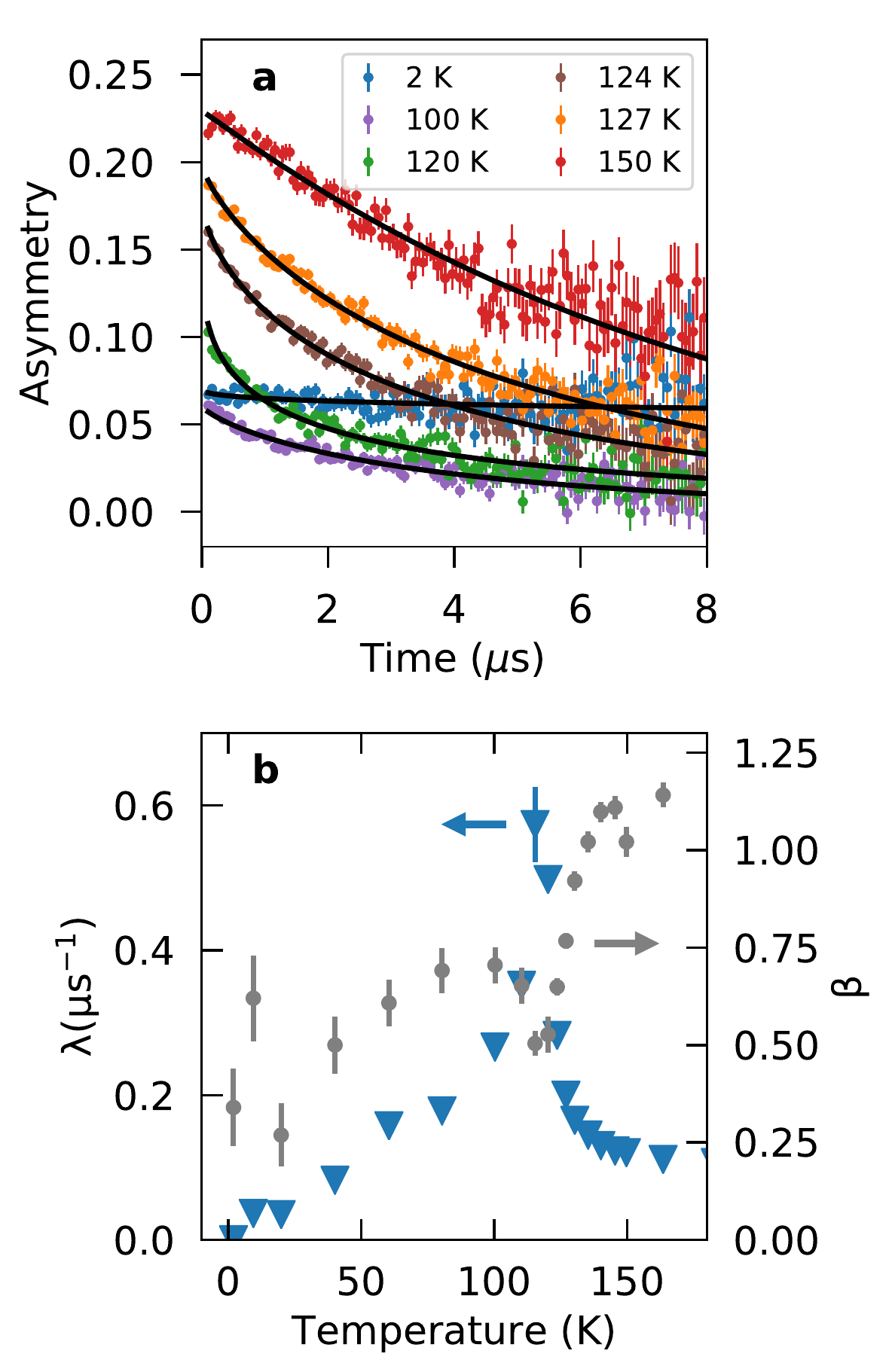}
	\caption{\label{fig:ZFlong} (a) Asymmetry spectra at representative temperatures above, below, and throughout the antiferromagnetic transition. Solid black curves are fits to the data using the model $a(t) = a_0 \exp (-\lambda t)^{\beta}$. (b) Temperature dependence of the relaxation rate $\lambda$ (blue triangles, left axis) and exponential power $\beta$ (gray circles, right axis) determined from the fits. The peak in $\lambda$ is characteristic of critical spin dynamics near a magnetic phase transition.}
		
\end{figure}

We first present the results of zero-field (ZF) \muSR\ experiments on MgO-ESO. Fig.~\ref{fig:ZFlong}(a) displays the time-dependent asymmetry $a(t)$ over 8~$\mu$s for several representative temperatures.
At 150~K, the initial asymmetry at $t = 0$ assumes the maximum observed value of $\sim$0.23 and then relaxes with time due to rapid fluctuations of the internal field in the paramagnetic phase. Two changes occur as the temperature is lowered: the initial asymmetry decreases, and the rate of the subsequent relaxation of the asymmetry becomes more rapid until about 120~K, below which the relaxation becomes slower once again. The decrease in initial asymmetry is caused by an extremely rapid depolarization process that appears as an instantaneous drop in asymmetry in the time binning used for Fig.~\ref{fig:ZFlong}(a), although closer inspection (to be shown later) reveals a continuous evolution of the asymmetry from its starting value. This behavior is due to the development of large and static internal fields as the AF transition is traversed. The evolution of the relaxation rate as the temperature is lowered is characteristic of changes in the spin dynamics in the vicinity of the AF transition.
	
Taking a more quantitative approach, we modeled each asymmetry spectrum with a ``stretched exponential'' function given by $a(t) = a_0 \exp(-\lambda t)^{\beta}$, where $a_0$ is the initial asymmetry amplitude, $\lambda$ is the relaxation rate, and $\beta$ is the power of the stretched exponential. In systems for which only one relaxation channel with a single well-defined spin fluctuation rate exists, $\beta$ is expected to be 1 and the relaxation follows a simple exponential function~\cite{uemur;prb85}. We found that such a model was inadequate to fit the observed spectra (particularly below the transition). Likewise, models using two or more exponential components were not stable and yielded fits with marginal quality in some temperature ranges. As has been done elsewhere~\cite{rover;prb02,dunsi;prb96}, we therefore used the phenomenological stretched exponential model. This indicates the presence of multiple relaxation channels and/or spin fluctuation rates, which is not surprising, given the various magnetic and nonmagnetic species present. The fits using the stretched exponential model are shown as the black curves in Fig.~\ref{fig:ZFlong}.
	
In Fig.~\ref{fig:ZFlong}(b), we display the temperature dependence of the refined relaxation rate $\lambda$ as blue triangles (left vertical axis) and the exponential power $\beta$ as gray circles (right vertical axis). The most striking feature in the relaxation rate is the presence of a pronounced peak with a maximum between 116 and 120~K. Such a peak is a hallmark of critical spin dynamics often associated with a continuous magnetic transition~\cite{reoti;jpcm97,uemur;ms99,goko;npjqm17}. As the transition is approached from above, the spin fluctuations undergo a ``critical slowing down'', leading to more efficient relaxation of the muon spin polarization and hence a divergence in the relaxation rate at the transition temperature. Below the transition, the relaxation is due to magnetic excitations that flip the muon spins. With decreasing temperature, fewer such excitations occur and the relaxation rate correspondingly decreases once more, until the asymmetry spectrum is essentially flat at 2~K. This indicates an absence of magnetic fluctuations at sufficiently low temperatures. The exponential power $\beta$ is near 1 in the paramagnetic phase but decreases to approximately 0.3 - 0.6 in the AF phase. Such behavior is not unusual for systems with significant randomness and disorder~\cite{suzuk;prb09}.
	
\begin{figure}
	\includegraphics[width=70mm]{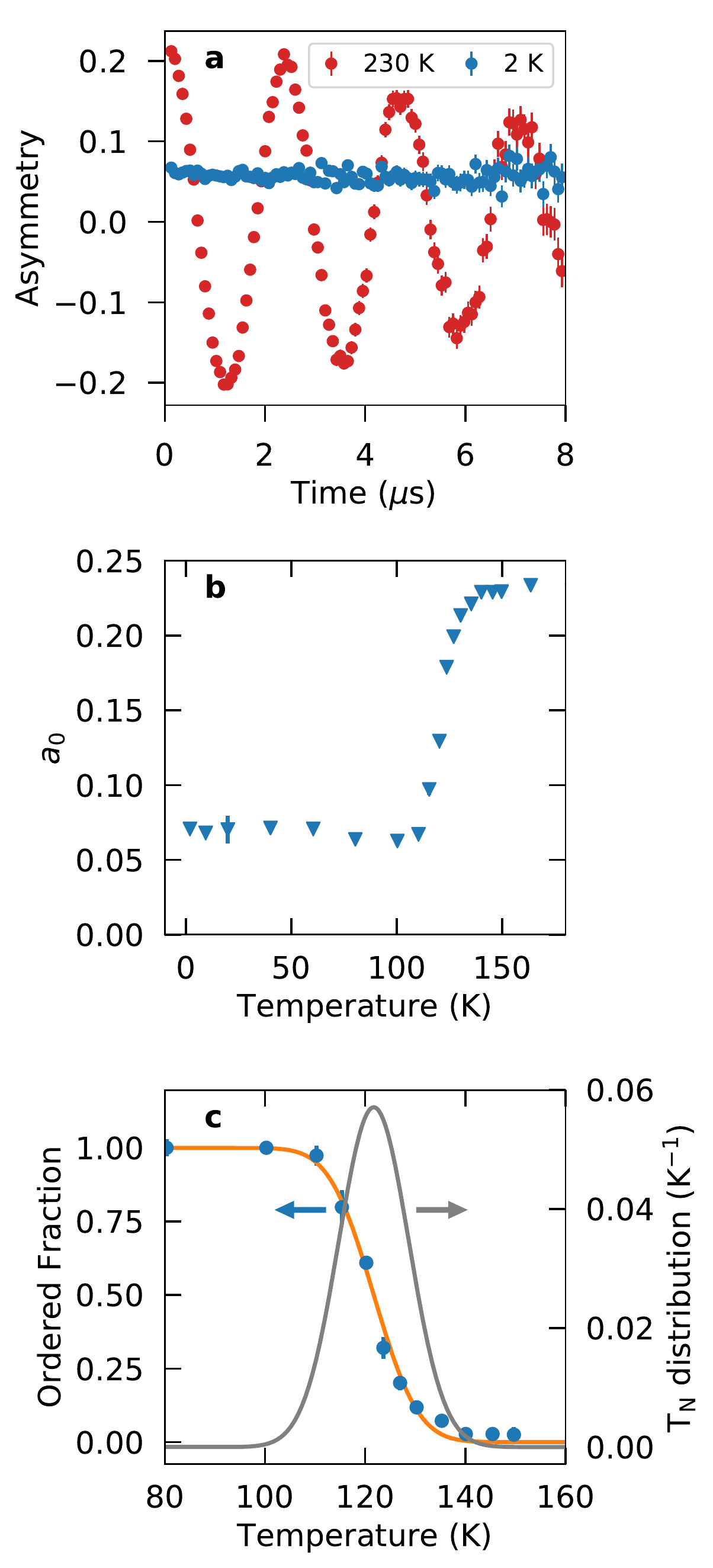}
	\caption{\label{fig:frac} (a) Weak transverse field (wTF) asymmetry spectra at 230~K and 2~K. The large-amplitude oscillations at high temperature indicate a fully paramagnetic state, while the complete lack of oscillations at low temperature indicates the presence of static magnetism extending through the entire sample volume. (b) Temperature dependence of the zero-field (ZF) asymmetry amplitude $a_0$ determined from fits to the spectra. (c) Antiferromagnetically ordered volume fraction (blue circles, left axis) calculated from the $a_0$ values. The orange curve is a fit to the volume fraction data using a complementary error function, corresponding to the Gaussian distribution of N\'eel temperatures shown as the gray curve (right axis).}
		
\end{figure}

We can also extract information about the magnetically ordered volume fraction from the \muSR\ data. Measurements conducted in a weak external field directed transverse to the initial muon spin polarization (wTF configuration) at 230~K and 2~K are shown in Fig.~\ref{fig:frac}(a). The oscillations observed at high temperature have a frequency that corresponds to the externally applied field of $\sim$30~G, demonstrating that the muon spins exhibit a net rotation around this external field, while the fluctuating internal fields result in no precession. This is characteristic of wTF measurements in the paramagnetic phase. On the other hand, the wTF spectrum at 2~K shows no oscillations, indicating that all muons land in an environment in which the internal fields are significantly larger than the external field and static on the muon time scale ($\mu$s). Therefore, we can conclude that static magnetism extends through 100\% of the sample volume at 2 K, with no regions of the sample remaining paramagnetic.

Returning now to the ZF \muSR\ data, we can obtain more detailed information about the temperature evolution of the magnetically ordered volume fraction. Fig.~\ref{fig:frac}(b) displays the temperature dependence of the ZF asymmetry amplitude $a_0$ determined from the stretched exponential fits. The continuous decrease in $a_0$ as the temperature is lowered from about 140~K to 100~K indicates that static magnetic order gradually develops throughout the sample across this temperature window. The remaining amplitude of $a_0 \sim 0.07$ at low temperature reflects the ``1/3 tail'' expected for a magnetically ordered polycrystalline sample, for which the internal magnetic field at the muon site in 1/3 of the crystallites on average will be along the initial muon spin polarization direction, thus preserving the polarization~\cite{uemur;ms99}. The magnetically ordered volume fraction $f$ can be calculated~\cite{frand;nc16} from the temperature-dependent $a_0 (T)$ values as $f(T) = [a_0^{max}-a_0(T)]/[a_0^{max}-a_0^{min}]$, where $a_0^{max}$ and $a_0^{min}$ are the maximum and minimum values of $a_0$, found at high and low temperature, respectively. The ordered fraction determined this way is shown by the blue circles in Fig.~\ref{fig:frac}(c). $f$ evolves smoothly between 100 and 140~K, confirming a broad transition during which phase separation between paramagnetic and antiferromagnetically ordered regions occurs.
	
The gradual development of the AF order over a wide temperature interval can be explained by the existence of a distribution of \TN\ values throughout the sample. The curvature of $f(T)$ closely resembles that of the complementary error function, which would result from a Gaussian distribution of ordering temperatures. Indeed, fitting the complementary error function to the $f(T)$ data results in the orange curve in Fig.~\ref{fig:frac}(c), which describes the data very well. The corresponding Gaussian distribution is shown as the gray curve in Fig.~\ref{fig:frac}(c), with a center of $T_0 = 121.5 \pm 0.5$~K and a standard deviation of $7.0 \pm 0.7$~K (FWHM = $16.5 \pm 1.6$~K). This distribution of \TN\ is therefore the most appropriate way to refer to the transition range of MgO-ESO, rather than just quoting a single value for the ordering temperature. In contrast to the broad transition observed here, many other conventional magnetic systems exhibit sharp transitions occurring over narrower temperatures ranges, often on the order of 1~K or less. We note that the center of the transition in MgO-ESO, $T_0 = 121.5 \pm 0.5$~K, corresponds closely to the maximum of the peak in the ZF relaxation rate.

\begin{figure*}
	\includegraphics[width=160mm]{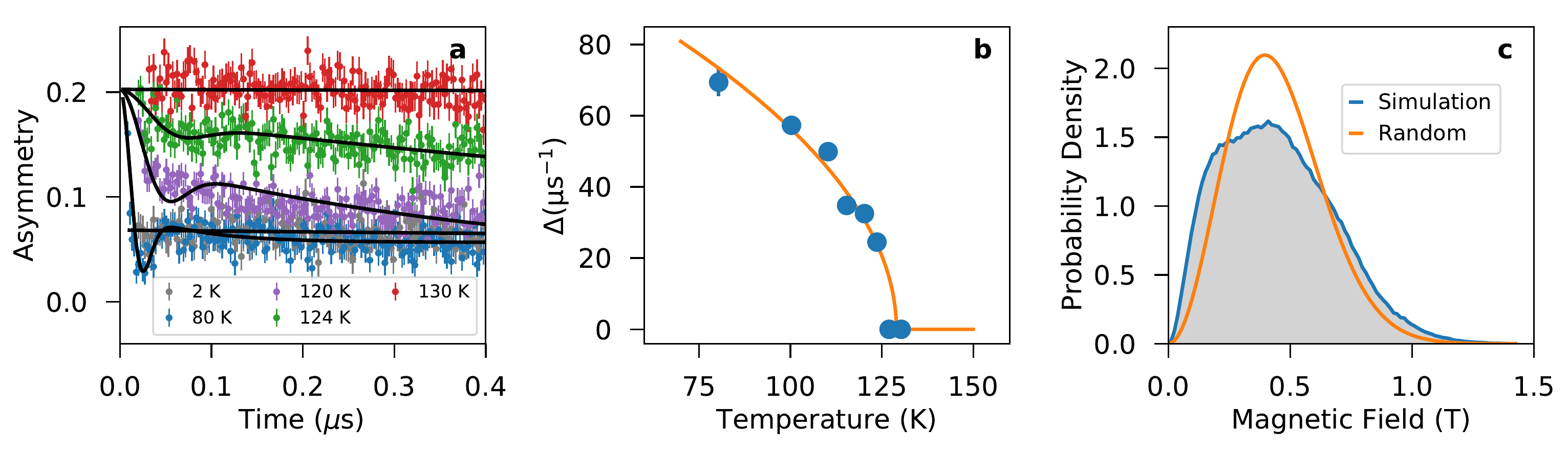}
	\caption{\label{fig:ZFearly} (a) Early-time portion of the zero-field \muSR\ asymmetry spectra for MgO-ESO at representative temperatures. The solid black curves represent fits using a model described in the main text. (b) The Kubo-Toyabe relaxation rate $\Delta$ (blue symbols) as a function of temperature extracted from the fits. The orange curve is a power law fit. (c) Simulated probability distribution of the magnetic field magnitude at a possible muon stopping site (blue curve), compared to the field distribution for a collection of completely randomly oriented spins corresponding to a KT distribution with $\Delta=210$~$\mu$s$^{-1}$ (orange curve).}
		
\end{figure*}

Further insight into the nature of the AF transition can be gained by examining the early-time portion of the ZF asymmetry spectra. Representative spectra are displayed in Fig.~\ref{fig:ZFearly}(a).
With this finer time binning, the continuous evolution of the asymmetry from the initial amplitude at $t = 0$ is evident. Particularly notable is the dip and partial recovery of the asymmetry, which is most obvious for the 80~K spectrum around 0.02~$\mu$s. This relaxation pattern is known as the Kubo-Toyabe (KT) relaxation function and is characteristic of static magnetic moments with significant disorder, resulting in a broad distribution of internal field values at the muon stopping sites~\cite{uemur;ms99}. The KT function is given by
\begin{align}\label{KT}
	G_{KT}(t)=\frac{1}{3}+\frac{2}{3}(1-\Delta^2t^2)\exp(-\Delta^2t^2/2)
\end{align}
where $\Delta$ is related to the width of the component-wise static internal field distribution $B_i$ through $\Delta / \gamma_{\mu} = B_i$, with $\gamma_{\mu} = 851.6$~$\mu$s$^{-1}$T$^{-1}$ the muon's gyromagnetic ratio. The early-time ZF spectra (up to 0.4~$\mu$s) can be fit adequately using a damped Kubo-Toyabe function and a simple exponential:	
\begin{align}\label{ZFearlyFunc}
	a(t) = a_{max}\left[ f_{KT}G_{KT}\exp(-\lambda_{KT}t) + (1-f_{KT})\exp(-\lambda t) \right],
\end{align}
where $a_{max}$ is the maximum initial asymmetry, $f_{KT}$ is the fraction of muons exhibiting KT relaxation, and $\lambda_{KT}$ and $\lambda$ are the relaxation rates in the ordered state and paramagnetic state, respectively. Fits using this model are shown as the black curves in Fig.~\ref{fig:ZFearly}(a). At 80~K, $\Delta$ is approximately 70~$\mu$s$^{-1}$, corresponding to a distribution of internal fields at the muon sites with a standard deviation of roughly 0.08~T. Below 80~K, the initial relaxation and partial recovery occur too rapidly to be observed within the time resolution of the data, preventing a precise measurement of $\Delta$. This is evident in the 2~K spectrum shown in Fig.~\ref{fig:ZFearly}(a). However, extrapolating the visible trend to zero temperature, $\Delta$ has an estimated value of 120-150~$\mu$s$^{-1}$, equivalent to a field distribution with a standard deviation of approximately 0.14-0.18~T.

Since the KT function is generally associated with random spin systems such as spin glasses, such a relaxation pattern may seem inconsistent with the reported long-range AF order in MgO-ESO. To investigate this, we performed dipole field calculations to simulate the field distribution of the published long-range AF structure at possible muon stopping sites. Following reports in the literature for simple binary oxides~\cite{uemur;hfi84,nishi;hfi97}, we chose muon stopping sites located 1~\AA\ away from the oxygen sites along the (111)-type directions in the cubic structure. For each assumed muon stopping site, we calculated the net magnetic field for 10$^4$ randomly generated cation configurations with the AF spin order imposed and the sublattice magnetization within the (111) plane (i.e., equivalent to the AF structure of NiO~\cite{roth;pr58}). The dipole sum included contributions from all spins within 10~\AA\ of the stopping site, which was sufficient to ensure reliable convergence. We assumed classical spin vectors with magnitudes of 1~\muB, 2~\muB, and 3~\muB\ for Cu$^{2+}$, Ni$^{2+}$, and Co$^{2+}$, respectively (and zero spin for Mg$^{2+}$ and Zn$^{2+}$). We then combined the results of these calculations for the 8 equivalent (111) directions.

This procedure yielded the distribution of magnetic field magnitudes shown by the blue curve and gray shaded area in Fig.~\ref{fig:ZFearly}(c). Due to the intrinsic randomness on the cationic sites, the field distribution is broad and qualitatively similar to that from a completely random spin configuration, despite the well-defined long-range AF order. The orange curve in Fig.~\ref{fig:ZFearly}(c) is the equivalent curve for a completely random magnetic field distribution with a component-wise standard deviation of 0.25~T. Such a distribution would give rise to an ideal KT relaxation function with $\Delta=210$~$\mu$s$^{-1}$, which agrees with the experimental results extrapolated to zero temperature ($\Delta\sim120-150$~$\mu$s$^{-1}$) to within a factor of order unity. Attempting to establish quantitative agreement is a futile endeavor without more accurate knowledge of the precise muon stopping site(s). However, the qualitative similarity seen in Fig.~\ref{fig:ZFearly}(c) between the simulated field distribution at a likely muon stopping site and a random distribution corresponding to ideal KT relaxation is sufficient to establish that the observed \muSR\ spectra need not be contradictory to the reported long-range AF order. 
	
To characterize the magnetic transition further, we use the parameter $\Delta$ in the KT function as an order parameter, since it is related to the magnitude of the static internal field. The temperature dependence of $\Delta$ is displayed as blue circles in Fig.~\ref{fig:ZFearly}(b), revealing a fairly continuous increase from zero as the temperature is lowered below $\sim$125~K. This is consistent with a continuous AF transition in MgO-ESO, as has also been observed in the binary oxides NiO and CoO~\cite{nishi;hfi97,srini;prb83,chatt;prb09}. The orange curve represents a power law fit using the equation $\Delta (T) = \Delta_0 \left( \frac{T_N - T}{T_N}\right)^{\beta}$, resulting in $T_N = 129 \pm 6$~K and $\beta = 0.51 \pm 0.17$. We stress once again that this best-fit value of $T_N$ should not be considered to be the ``true'' N\'eel temperature, since in actuality there is a distribution of N\'eel temperatures. We also acknowledge that the value of the critical exponent $\beta$ has a large uncertainty due to a limited number of data points and the complications associated with the broad distribution of ordering temperatures. Nevertheless, these values may be useful for comparison with other techniques. 
	
\begin{figure}
	\includegraphics[width=70mm]{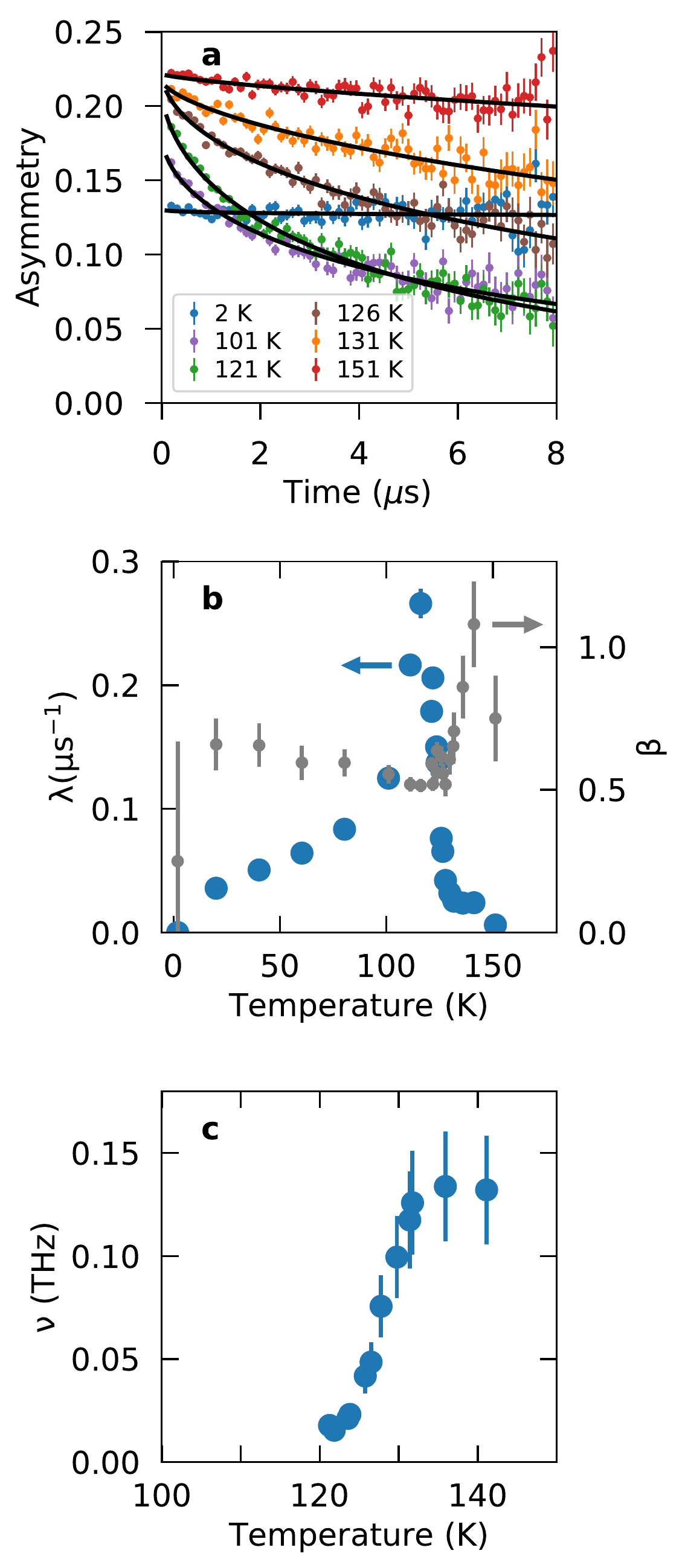}
	\caption{\label{fig:LFfull} (a) Asymmetry spectra in longitudinal field (LF) of 0.3~T shown for representative temperatures. The solid black curves are fits using the model $a(t) = a_0 \exp (-\lambda t)^{\beta}$. (b) Temperature dependence of the relaxation rate $\lambda$ (blue circles, left axis) and exponential power $\beta$ (gray circles, right axis) determined from the fits. (c) Estimated spin fluctuation rate $\nu$ determined from $\lambda$ as described in the text.
}		
\end{figure}

To supplement the ZF \muSR\ results, we also performed longitudinal-field (LF) measurements, in which a magnetic field was applied to the sample parallel to the initial muon spin polarization direction. Asymmetry spectra collected in a longitudinal field of 0.3~T at representative temperatures are displayed in Fig.~\ref{fig:LFfull}(a).
As with the ZF spectra, the spectra can be well described by a stretched exponential function of the form $a(t) = a_0\exp(-\lambda t)^{\beta}$. The fits are shown as the black curves overlaid on the data in Fig.~\ref{fig:LFfull}(a). The extracted values of $\lambda$ and $\beta$ are plotted as a function of temperature in Fig.~\ref{fig:LFfull}(b) using blue circles (left axis) and smaller gray circles (right axis), respectively. Both parameters show a very similar temperature dependence to that of the ZF data. In particular, the LF relaxation rate also exhibits a dramatic peak centered around 120~K, confirming the presence of critical dynamics near the transition.

In a system with a single relaxation channel and spin fluctuation rate $\nu$ in the paramagnetic phase, the ZF or LF relaxation rate $\lambda$ can be related to $\nu$ according to the formula~\cite{uemur;prb85}
\begin{align}\label{lambda-fluctuation}
	\lambda = \frac{2\Delta^2\nu}{\nu^2+\omega_L^2},
\end{align}
where $\Delta = \gamma_{\mu}B_i$ is the product of the rms internal magnetic field $B_i$ and the muon's gyromagnetic ratio, and $\omega_L = \gamma_{\mu}B_{L}$ is the Larmor frequency of the muon spin in the longitudinally applied field $B_L$. Inverting this equation to solve for $\nu$ yields 
\begin{align}\label{fluctuation-lambda}
	\nu = \frac{\Delta^2+\sqrt{\Delta^4-\lambda^2\omega_L^2}}{\lambda}.
\end{align}
This allows the calculation of $\nu$ from the experimentally determined values of $\Delta, \lambda,$ and $\omega_L$. In the present case, a simple exponential function is not sufficient to describe the observed relaxation spectra in ZF or LF, meaning that the assumption of a single relaxation channel and spin fluctuation rate does not strictly hold and this formula for $\nu$ is not rigorously correct. Nevertheless, we can use this formula to provide a reasonable estimate of $\nu$ as a function of temperature. Because some of the ZF relaxation arises from weak nuclear dipolar fields of the atoms, we consider the LF relaxation rates to be more reliable for estimating $\nu$. As has been done elsewhere~\cite{dunsi;prb96}, $\Delta$ can be estimated from the low-temperature ZF data, leading to a value of 100~$\pm$~10~$\mu$s$^{-1}$ in the present case. The value of $\omega_L$ is set by the external field of 0.3~T. Fig.~\ref{fig:LFfull}(c) shows the resulting estimated values of $\nu$ determined according to Eq.~\ref{fluctuation-lambda} using the LF relaxation rates on the high-temperature side of the peak in Fig.~\ref{fig:LFfull}(b). The rapid decrease of $\nu$ between 140~K and 120~K reflects the critical slowing down of the spin fluctuations as the transition is approached. We note that $\nu$ calculated from the ZF data (not shown) agrees qualitatively with the LF results at all temperatures and quantitatively between 120 and 130~K, where any effects of the nuclear dipolar fields are completely masked by the effects of the electronic spins.

\subsection{Inelastic Neutron Scattering and Thermodynamic Properties}
\begin{figure*}
	\includegraphics[width=120mm]{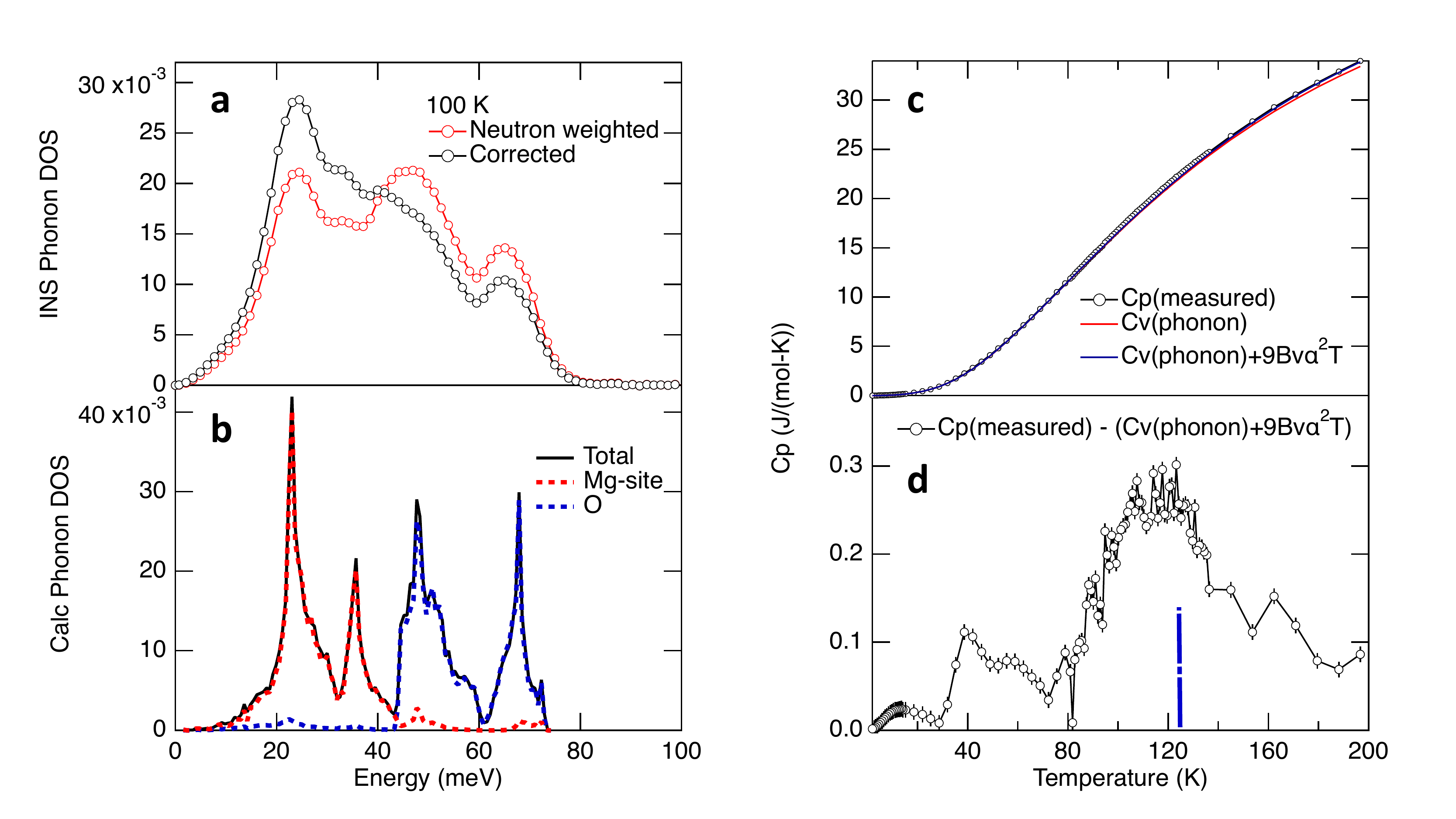}
	\caption{\label{fig:phonon} Thermodynamic analysis using phonon properties determined from inelastic neutron scattering (INS) measurements combined with calorimetry. (a) Phonon density of states (DOS) extracted from INS. The red points give the DOS extracted from the measured scattering function before correcting for atomic cross sections and masses. The black points give the thermodynamic phonon DOS determined by correcting for cross sections and atomic contributions based on calculations of the phonon DOS shown in (b). The total calculated phonon DOS (black line) is well separated into Mg-site (red) and O (blue) contributions. (c) Measured heat capacity, $C_P$ (black points), along with the phonon contribution at constant volume ($C_V$) determined from the measured thermodynamic phonon DOS. A correction for volume expansion ($9Bv\alpha^2 T$) is also included. (d) Magnetic contribution determined by subtracting the phonon contribution from the measured heat capacity. A very broad magnetic transition is evident.
	}		
\end{figure*}
We now provide additional analysis of the inelastic neutron scattering (INS) and thermodynamic data reported in our earlier work~\cite{zhang;cm19}, in the context of the broad transition established by the \muSR\ results. The specific heat reported in Ref.~\onlinecite{zhang;cm19} does not display a prominent feature indicative of any long-range magnetic transition, which seems somewhat puzzling. We can examine this further by estimating the specific heat jump expected at \TN\ using the experimental magnetic DC- and AC-susceptibilities~\cite{zhang;cm19} and the method in Ref.~\onlinecite{fishe;pm62}. This estimate indicates that we should observe a decrease of $\sim$1.25~J/mol/K stretched over $\sim$20~K across \TN. Such a change is not visible in the specific heat data of Ref.~\onlinecite{zhang;cm19}. However, a better assessment of the magnetic specific heat, \CM, can be obtained by subtracting the phonon contribution to the specific heat from the experimental data. To determine the phonon contribution, we make use of the INS data reported in Ref.~\onlinecite{zhang;cm19} using the ARCS time-of-flight spectrometer~\cite{abern;rsi12} at the Spallation Neutron Source (SNS) of Oak Ridge National Laboratory. Measurements were performed at 100~K and 300~K using 140-meV incident energy neutrons. In order to obtain a good average over the Brillouin zone and to avoid magnetic scattering at low momentum transfer, $Q$, the spectrum was integrated from $Q$ = 6~\AA$^{-1}$ to 15~\AA$^{-1}$ and corrected for multiphonon scattering, elastic scattering, and thermal occupation using a procedure described elsewhere~\cite{manle;prb02}. 

Fig.~\ref{fig:phonon}(a) shows the resulting neutron weighted phonon density of states (DOS). The initially determined phonon DOS is neutron weighted because inelastic scattering intensities depend on the atomic cross sections divided by atomic masses. For this reason, the average scattering intensity from the O site is about 1.75 times larger than the average scattering from the Mg-site (average of Mg, Co, Ni, Cu, and Zn atoms). This can only be corrected with a calculation to determine the atomic partial contributions to the phonon DOS. Fig.~\ref{fig:phonon}(b) shows a calculated total phonon DOS with the breakdown for the Mg-site and O-site contributions. This calculation is based on DFT-calculated forces from MgO~\cite{parli;10}, but with the Mg mass changed to reflect the average Mg-site mass for MgO-ESO. The calculation shows that the partial contributions are well split between the low-energy modes being dominated by the Mg-site atoms and the higher-energy modes being dominated by the O atoms. Accounting for the weighted contributions, we have corrected the neutron weighted phonon DOS, shown as the corrected curve in Fig.~\ref{fig:phonon}(a), which tends to distribute more density to lower energies. The corrected phonon DOS was then used to calculate the constant volume phonon contribution to the heat capacity (per atom) using, 
\begin{align}\label{CM}
	C_{V,phonon}(T)=3k_B\int_{0}^{\infty}g(\omega)\left(\frac{\hbar\omega}{k_BT}\right)^2\frac{e^{\hbar\omega / k_B T}}{e^{\hbar \omega / k_B T}-1}\mathrm{d}\omega,
\end{align}
where $g(\omega)$ is the corrected phonon DOS. Fig.~\ref{fig:phonon}(c) shows a comparison of this calculated phonon contribution and the measured heat capacity at constant pressure, $C_P$. A small additional correction for thermal expansion was added using the thermodynamic identity $C_P-C_V=9Bv\alpha^2 T$, where $v$ is the specific volume, $\alpha$ is the linear thermal expansion coefficient determined from the temperature dependence of the measured lattice parameter~\cite{zhang;cm19}, and $B\sim185$~GPa is the bulk modulus obtained by density functional theory calculations~\cite{pitik;jap20}. 

After performing these calculations, the magnetic contribution to the specific heat, obtained as $C_M  = C_P - C_{V,phonon} – 9Bv\alpha^2 T$, is found to display a broad peak across the transition centered around $\sim$120 K, as shown in Fig.\ref{fig:phonon}(d). The features of this peak are in good agreement with the gradual transition obtained from the \muSR\ data, but still smaller (by a factor of 4-5) than expected from the estimation based on the magnetic susceptibility. The failure of the simple model by Fisher~\cite{fishe;pm62} likely originates from the very broad transition and the presence of three different spin values for Co, Ni, and Cu. In addition, the magnetic entropy release obtained from integrating $C_M/T$ between 72~K and 196~K is 0.156~J~(mol~K)$^{-1}$, significantly less than the maximum value that would be expected for full ordering of this system. The temperature range for integration was selected because it spans the broad hump in $C_M$ associated with the transition. The discrepancy is likely due to the ordered moment never reaching full saturation and/or additional undetected entropy release at higher temperature~\cite{zhang;cm19}. A more complete understanding of the thermodynamics will require future detailed modeling of the spin-waves in these entropy-stabilized oxides.

\begin{figure}
	\includegraphics[width=70mm]{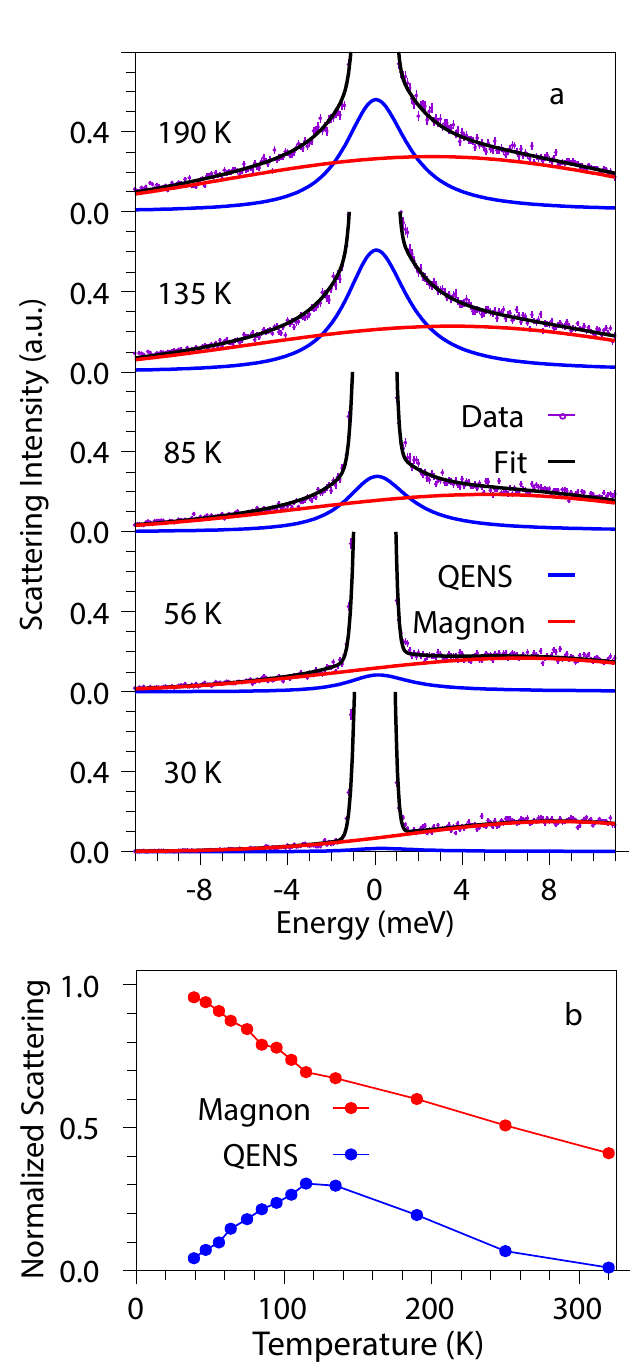}
	\caption{\label{fig:INS} (a) Inelastic neutron scattering (INS) spectra for MgO-ESO at representative temperatures obtained from integrating over the $Q$-range $1.1 < Q < 1.4$~\AA$^{-1}$. The solid curves represent fits described in the main text. (b) Normalized weights of the magnon and quasielastic neutron scattering (QENS) components of the fits to the INS spectra as a function of temperature. See main text for additional details.
	}		
\end{figure}

The high-resolution INS data collected on HYSPEC provide additional information about the spin dynamics in MgO-ESO below the ordering temperature. Fig.~\ref{fig:INS}(a) displays several INS spectra collected at representative temperatures, obtained through integration over the range $1.1 < Q < 1.4$~\AA$^{-1}$. This $Q$ range was selected because it includes the first antiferromagnetic Bragg peak centered around 1.27~\AA$^{-1}$~\cite{zhang;cm19}.
A magnon gap of approximately 7~meV is observed in the data collected between 5 and 30~K, as seen by the suppression of scattered intensity above the elastic bandwidth of $\sim$1~meV. With increasing temperature, the intensity at low energy transfers increases, indicating that the magnon gap is gradually covered by quasielastic spin relaxation scattering. To analyze the data, we have fitted the signal at 30~K with a phenomonological Gaussian magnon contribution (energy of 4.5~meV; FWHM of 17~meV) and the elastic scattering resolution function. Then, for all subsequent data at higher temperature, we added a Lorentzian quasielastic scattering (QENS) signal and varied only the relative weight of magnon and QENS signals, taking into account the thermal population and detailed balance~\cite{zheng;sadv19} at each specific temperature. In initial fits, the QENS signal indicated an essentially constant Lorentzian width of $\sim$3.5~meV, which we subsequently fixed for the final series of fits, displayed in Fig.~\ref{fig:INS}(a). We observe a gradual increase in the QENS intensity at the expense of the magnon scattering upon heating until \TN\ is reached, as seen in Fig.~\ref{fig:INS}(b). Above \TN, both the QENS and magnon scattering decrease with increasing temperature, presumably because the signal moves out of the available energy window of $\pm$11~meV. The observed QENS width of 3.5(5)~meV FWHM indicates that the magnetic excitations have a lifetime of approximately 190(30)~fs. 

The temperature independence of the QENS width is unexpected and suggests some sort of intrinsic fluctuation time scale, in analogy to the Kondo problem. Here however, as the material is an insulator, the scattering rate cannot result from the hybridization of localized electrons with the conduction band. These intrinsic fluctuations are instead likely to be related to the magnetic site disorder. Further investigation of their origin is beyond the scope of this study. The gradual suppression of spin excitations as the temperature is lowered is qualitatively consistent with the gradual decrease in the \muSR\ relaxation rate $\lambda$ observed in Fig.~\ref{fig:ZFlong}(b) and Fig.~\ref{fig:LFfull}(b), although we note that \muSR\ and INS probe spin excitations on different energy and time scales.

\section{Discussion and Conclusion}	
The results reported here provide several new insights into the nature of the AF state in MgO-ESO. First, the \muSR\ measurements demonstrate unambiguously that the magnetic order extends through the entire sample volume, excluding the possibility of any large regions that remain magnetically disordered at low temperature. Second, we have shown that the AF state develops gradually throughout the sample volume between 100 and 140~K, with the evolution of the ordered volume fraction being well described by a Gaussian distribution of ordering temperatures centered at 121.5~$\pm$0.5~K with a standard deviation of 7.0~$\pm$~0.7~K. This is consistent with the broad peak in \CM\ revealed by analysis of the specific heat using the INS data. Third, the temperature dependence of the ZF and LF \muSR\ relaxation rates exhibits a prominent peak around 120~K due to critical spin dynamics in the vicinity of the transition, and the static internal field at the muon site grows continuously from 0 as the transition is traversed. Finally, excitations in the ordered state are frozen out at low temperature due to the formation of a magnon gap of $\sim$7~meV.

These findings help address many of the open questions left after the first studies of MgO-ESO~\cite{jimen;apl19,zhang;cm19}. The discrepancy between the onset of magnetic Bragg peak intensity at 140~K, the peak in the magnetic susceptibility at 113~K, and the lack of any apparent signature of the transition in the specific heat as reported originally in Ref.~\onlinecite{zhang;cm19} can be explained naturally by the gradual development of AF order in the sample between 100 and 140~K. As soon as the first regions of the sample order around 140~K, magnetic Bragg peak intensity will appear in the neutron diffraction pattern. Around that same temperature, the reciprocal magnetic susceptibility starts to deviate from the high temperature linear temperature dependence~\cite{zhang;cm19}, consistent with the development of AF order in a partial volume fraction. A peak in the temperature-dependent magnetic susceptibility is observed once a large enough fraction of the sample has ordered (and therefore has vanishing magnetization for an AF system), explaining the presence of the susceptibility peak at the lower temperature of 113 K. The entropy release from the AF transition is significantly broadened due to the large temperature width of the transition and the presence of strong fluctuations, so the lack of an obvious signature in the specific heat data shown in Ref.~\onlinecite{zhang;cm19} is not particularly surprising. Only after carefully isolating the magnetic contribution to the specific heat, as accomplished in the present work, does a broad feature in the vicinity of the transition appear.

Regarding the reduction of the ordered moment from the average value of 2~\muB\ expected for fully ordered Cu$^{2+}$, Ni$^{2+}$, and Co$^{2+}$ moments to the observed value of 1.4~\muB, explanations offered previously~\cite{zhang;cm19} include poorly connected magnetic moments that do not order, residual magnetic fluctuations, covalency with oxygen atoms, or some other influence of the chemical disorder. From the \muSR\ data presented here, the first two scenarios can be ruled out, since we observe no significant regions that remain disordered below 100~K, and the ZF and LF spectra at base temperature show no relaxation, indicating that no significant fluctuations are present at such low temperatures. This is consistent with non-relaxing spectra at low temperature in NiO and CoO~\cite{nishi;hfi97}. Therefore, covalency with the oxygen atoms or some other effect of the chemical disorder are the most likely explanation for the reduced moment. However, it is still unclear whether the broad AF transition originates in covalency with oxygen or chemical disorder rather than more simply the dilution of the spins and disorder in their magnitude. Detailed calculations of how the percolation of the magnetic order proceeds are required to settle this question.

Finally, the peaks in the ZF and LF relaxation rates and the continuous growth of the static internal field revealed by the \muSR\ experiments provide the first clear evidence that the character of the AF transition in MgO-ESO is essentially continuous, albeit distributed over a broad temperature interval. That the transition is continuous rather than first order in the presence of such extreme disorder may seem somewhat surprising. However, disorder need not be detrimental to a continuous phase transition; in fact, disorder can sometimes restore a continuous transition that is otherwise hidden by a first-order transition, as has been shown recently for the helimagnet MnSi~\cite{goko;npjqm17}. In the present case, the continuous nature of the AF transition is similar to the continuous AF transitions in CoO and NiO~\cite{nishi;hfi97,srini;prb83,chatt;prb09}, but in contrast to the first-order transition in MnO~\cite{uemur;hfi84}. In MnO, the AF transition is strongly coupled to a structural distortion that occurs simultaneously with the magnetic transition, causing the AF transition to become first order~\cite{uemur;hfi84}. On the other hand, the structural distortion in NiO is much more weakly coupled to the AF transition and occurs at a temperature below \TN, preserving the continuous nature of the AF transition in NiO~\cite{chatt;prb09,balag;jetpl16}. No such structural distortion has been observed at any temperature in MgO-ESO, perhaps due to the difficulty of establishing a coherent structural phase transition in the presence of such extreme local disorder. In this scenario, the lack of any structural phase transition may also help preserve the continuous nature of the AF transition in MgO-ESO. Further experimental and theoretical studies will be necessary to understand this issue more fully.
	
	
\textbf{Acknowledgements}
	
We thank Gerald Morris, Bassam Hitti, and Donald Arseneau for their support at the \muSR\ beamline at TRIUMF, John Budai and Doug Abernathy for their help at the ARCS beamline, and Melissa Graves-Brook for her help at the HYSPEC beamline. The work at Brigham Young University (BYU) was made possible by funds provided by the College of Physical and Mathematical Sciences at BYU. Material synthesis, inelastic neutron scattering and muon spin resonance work by RPH (INS, \muSR), MEM (INS) and JJZ (INS, synthesis), and JY (synthesis) was supported by the US Department of Energy, Office of Science, Office of Basic Energy Sciences, Materials Sciences and Engineering Division and work by BW (INS) was supported by was supported by the United States Department of Energy, Office of Science, Office of Basic Energy Sciences , Scientific User Facilities Division, both under Contract Number DE-AC05-00OR22725. This study used in parts resources at the Spallation Neutron Source (SNS), a DOE Office of Science User Facility operated by the Oak Ridge National Laboratory (INS measurements performed at the ARCS and HYSPEC beamlines).

%

\end{document}